\documentclass [11pt,a4paper] {article}

\usepackage[cp1252]{inputenc}
\usepackage{amssymb}
\usepackage{amsmath}
\usepackage{amsfonts,amssymb}
\usepackage[dvips]{graphicx}
\usepackage{bbm}
\usepackage{enumerate}
\usepackage{amsthm}
\usepackage{cancel}

\DeclareMathAlphabet{\mathpzc}{OT1}{pzc}{m}{it}

\newtheorem{definition}{Definition} 
\newtheorem{hypothesis}{Hypothesis} 

\begin{document}

\title{Propositional counter-factual definiteness and the EPR paradox}

\author{Arkady Bolotin\footnote{$Email: arkadyv@bgu.ac.il$\vspace{5pt}} \\ \textit{Ben-Gurion University of the Negev, Beersheba (Israel)}}

\maketitle

\begin{abstract}\noindent In an empirical logic, an experimentally verifiable proposition $P$ relating to a quantum system is assigned the value of either true of false if the system is in the pure state that belongs or, respectively, does not belong to the Hilbert subspace that represents $P$. Determined in such a way truth or falsity of $P$ can be termed \textit{a factual truth-value} of $P$. In the present paper, it is proposed to consider \textit{a counter-factual truth-value} of $P$, i.e., either of the values, true or false, that might have been taken by $P$ if the system had been in a pure state belonging to a Hilbert subspace that does not represent $P$. The assumption that it is always possible to speak meaningfully of counter-factual truth-values of experimental propositions can be called \textit{the hypothesis of propositional counter-factual definiteness}. As it is shown in the paper, this hypothesis lies at the basis of the EPR paradox, a striking and influential thought experiment intended to defy predictions of quantum mechanics, such as one that measurements of spin along the different axes are incompatible. The purpose of this paper is to show that this hypothesis can be falsified by declining to paste together invariant-subspace lattices of contexts associated with the system (in other words, Boolean algebras or blocks) into one Hilbert lattice. Without such pasting, the EPR paradoxical inference cannot be reached.\\

\end{abstract}

\section{Introduction}  

\noindent As stated by the axioms underlying an empirical logic \cite{Mackey, Putnam}, \textit{quantum propositions} – i.e., experimentally verifiable (or at least potentially falsifiable) propositions relating to a quantum system – are represented by corresponding closed linear subspaces of a Hilbert space $\mathcal{H}$ associated with the system. Following upon that, conditions and rules imposed on the quantum propositions are determined by partial binary lattice-theoretic operations on those subspaces \cite{Pavicic}.\\

\noindent To implement this logic, the quantum proposition $P$ is assigned the value of \textit{true} if the system is in the pure state $|\Psi\rangle$ belonging to the subspace $\mathcal{H}_P$ that represents $P$ \cite{Redei}. This can be expressed as\smallskip

\begin{equation} \label{RULE} 
   {\big[\mkern-4.3mu\big[ 
      P \left(
         |\Psi\rangle \in \mathcal{H}_P
      \right)
   \big]\mkern-4.3mu\big]}_v
   =
   \mathrm{true}
   \;\;\;\;  ,
\end{equation}
\smallskip

\noindent where $P(|\Psi\rangle \in \mathcal{H}_P)$ stands for ``$P$ in the state $|\Psi\rangle \in \mathcal{H}_P$''. Otherwise, i.e., if the system is in the pure state that does not reside in $\mathcal{H}_P$, the proposition $P$ is assigned the value of \textit{false}:\smallskip

\begin{equation}  
   {\big[\mkern-4.3mu\big[ 
      P \left(
         |\Psi\rangle \notin \mathcal{H}_P
      \right)
   \big]\mkern-4.3mu\big]}_v
   =
   \mathrm{false}
   \;\;\;\;  .
\end{equation}
\smallskip

\noindent Along with determined in such a way truth or falsity of the proposition $P$, which can be termed \textit{a factual truth-value} of $P$, one can consider \textit{a counter-factual truth-value} of $P$, namely, either of the values, true or false, that might have been taken by $P$ if, instead of the state $|\Psi\rangle \in \mathcal{H}_P$ or $|\Psi\rangle \notin \mathcal{H}_P$, the system had been in the state $|\Psi\rangle \in \mathcal{H}_Q$ where $\mathcal{H}_Q$ is a subspace of $\mathcal{H}$ \textit{that does not represent} $P$. Whether it is possible to treat a counter-factual truth-value as it were a factual one, that is, if it is possible to speak meaningfully of the truth-value of the quantum proposition $P$ in the state $|\Psi\rangle \in \mathcal{H}_Q$, must apparently depend on the lattice-theoretic \textit{ordering relation} $\le$ over the subspaces $\mathcal{H}_Q$ and $\mathcal{H}_P$.\\

\noindent For example, suppose that $\mathcal{H}_Q \le \mathcal{H}_P$; this means that $\mathcal{H}_Q$ is co-aligned with $\mathcal{H}_P$ such that $\mathcal{H}_Q$ is a subset of $\mathcal{H}_P$. In that case, the state $|\Psi\rangle$, which lies in the subspace $\mathcal{H}_Q$, belongs to the subspace $\mathcal{H}_P$ as well; therefore, the counter-factual truth-value of the quantum proposition $P$ in the state $|\Psi\rangle \in \mathcal{H}_Q$ can be treated the same as the factual truth-value of $P$ in the state $|\Psi\rangle \in \mathcal{H}_P$. In symbols,\smallskip

\begin{equation}  
   \mathcal{H}_Q
   \le
   \mathcal{H}_P
   \textnormal{:}
   \;\;
   {\big[\mkern-4.3mu\big[ 
      P \left(
         |\Psi\rangle \in \mathcal{H}_Q
      \right)
   \big]\mkern-4.3mu\big]}_v
    =
   {\big[\mkern-4.3mu\big[ 
      P \left(
         |\Psi\rangle \in \mathcal{H}_P
      \right)
   \big]\mkern-4.3mu\big]}_v
   =
   \mathrm{true}
   \;\;\;\;  .
\end{equation}
\smallskip

\noindent By contrast, suppose that $\mathcal{H}_Q \nleq \mathcal{H}_P$ and $\mathcal{H}_P \nleq \mathcal{H}_Q$. This means that the state $|\Psi\rangle \in \mathcal{H}_Q$ can be a member of $\mathcal{H}_P$ only if $|\Psi\rangle = 0$, where 0 is the zero-vector, i.e., the solitary vector contained in the zero-dimensional subspace $\{0\}$, a proper subset of any closed linear subspace of $\mathcal{H}$. But, since any state $|\Psi\rangle$, full of physical meaning, must differ from 0, it follows that the counter-factual truth-value of $P$ in $|\Psi\rangle \in \mathcal{H}_Q$ can be treated the same as the factual truth-value of $P$ in $|\Psi\rangle \notin \mathcal{H}_P$, namely,\smallskip

\begin{equation} \label{FALS} 
   \mathcal{H}_Q
   \nleq
   \mathcal{H}_P
   \textnormal{,}
   \,
   \mathcal{H}_P
   \nleq
   \mathcal{H}_Q
   \textnormal{:}
   \;\;
   {\big[\mkern-4.3mu\big[ 
      P \left(
         |\Psi\rangle \in \mathcal{H}_Q
      \right)
   \big]\mkern-4.3mu\big]}_v
    =
   {\big[\mkern-4.3mu\big[ 
      P \left(
         |\Psi\rangle \notin \mathcal{H}_P
      \right)
   \big]\mkern-4.3mu\big]}_v
   =
   \mathrm{false}
   \;\;\;\;  .
\end{equation}
\smallskip

\noindent The assumption that the counter-factual truth value can be definite for any pair of subspaces $\mathcal{H}_Q$ and $\mathcal{H}_P$, that is,\smallskip

\begin{equation} \label{HYP} 
   \forall
   \left\{
      \mathcal{H}_Q
      ,
      \mathcal{H}_P
   \right\}
   \textnormal{:}
   \;\;
   {\big[\mkern-4.3mu\big[ 
      P \left(
         |\Psi\rangle \in \mathcal{H}_Q
      \right)
   \big]\mkern-4.3mu\big]}_v
    =
   \mathfrak{t}
   \in
   \{ \mathrm{true},\mathrm{false} \}
   \;\;\;\;  ,
\end{equation}
\smallskip

\noindent can be called \textit{the hypothesis of propositional counter-factual definiteness}. The purpose of this paper is to show that this hypothesis can be falsified by declining to paste together invariant-subspace lattices of different contexts associated with a quantum system (i.e., Boolean algebras or blocks) into a Hilbert lattice of the system. Without such pasting, the counter-factual inference lying at the basis of the EPR paradox (arguing against the prediction of quantum mechanics maintaining that measurements of spin along the different axes are incompatible) cannot be reached.\\

\section{Definitions and preliminaries}  

\noindent Before demonstrating this, some definitions and preliminaries are in order first.\\

\begin{definition} 
Any closed linear subspace $\mathcal{H}_P$ of a Hilbert space $\mathcal{H}$ is the range of some projection operator $\hat{P}$ (i.e., self-adjoint idempotent operator) acting on $\mathcal{H}$ \textnormal{\cite{Kalmbach}}. Explicitly, $\mathcal{H}_P$ is identical to the subset of the vectors $|\Psi\rangle \in \mathcal{H}$ that are in the image of the projection operator $\hat{P}$\textnormal{:}\smallskip

\begin{equation} \label{RAN} 
   \mathcal{H}_P
   \equiv
   \mathrm{ran}(\hat{P})
   =
   \left\{
      |\Psi\rangle \in \mathcal{H}
      \textnormal{:}
      \mkern10mu
      \hat{P} |\Psi\rangle
      =
      |\Psi\rangle
   \right\}
   \;\;\;\;  .
\end{equation}
\smallskip

\end{definition}

\noindent In the same way, the closed linear subspace $\mathcal{H}_P^\perp \subseteq \mathcal{H}$, which is \textit{orthogonal} to $\mathcal{H}_P$, is identical to \textit{the kernel} of $\hat{P}$, i.e., the subset of the vectors $|\Psi\rangle \in \mathcal{H}$ that are mapped to zero by $\hat{P}$, explicitly,\\

\begin{equation} \label{KER} 
   \mathcal{H}_P^{\perp}
   \equiv
   \mathrm{ker}(\hat{P})
   =
   \mathrm{ran}(\hat{1} - \hat{P})
   =
   \left\{
      |\Psi\rangle \in \mathcal{H}
      \textnormal{:}
      \mkern10mu
      (\hat{1} - \hat{P}) |\Psi\rangle =  |\Psi\rangle
   \right\}
   \;\;\;\;  ,
\end{equation}
\smallskip

\noindent where $\hat{1}$ stands for the identity operator on $\mathcal{H}$. For that reason, the projection operator $\hat{1}-\hat{P}$ can be understood as the negation of $\hat{P}$, i.e.,\smallskip

\begin{equation}  
   \neg\hat{P}
   =
   \hat{1} - \hat{P}
   \;\;\;\;  .
\end{equation}
\smallskip

\noindent As consequences of (\ref{RAN}) and (\ref{KER}), one has\smallskip

\begin{equation}  
   \mathcal{H}_P
   \cap
   \mathcal{H}_P^{\perp}
   =
   \mathrm{ran}(\hat{P})
   \cap
   \mathrm{ran}(\neg\hat{P})
   =
   \mathrm{ran}(\hat{0})
   =
   \{0\}
   \;\;\;\;  ,
\end{equation}

\begin{equation}  
   \mathcal{H}_P
   +
   \mathcal{H}_P^{\perp}
   =
   \mathrm{ran}(\hat{P})
   +
   \mathrm{ran}(\neg\hat{P})
   =
   \mathrm{ran}(\hat{1})
   =
   \mathcal{H}
   \;\;\;\;  ,
\end{equation}
\smallskip

\noindent where $\cap$ denotes the set-theoretic operation of intersection, $\hat{0}$ is the zero operator on $\mathcal{H}$, the sum of the subspaces $\mathcal{H}_P + \mathcal{H}_P^{\perp}$ is defined as
the set of sums of vectors from $\mathcal{H}_P$ and $\mathcal{H}_P^{\perp}$, while the subsets $\{0\}$ and $\mathcal{H}$ are \textit{the trivial subspaces of $\mathcal{H}$} (which correspond to the trivial projection operators $\hat{0}$ and $\hat{1}$, respectively).\\

\begin{definition} 
A set of two or more nontrivial projection operators $\hat{P}_A$, $\hat{P}_B$, $\dots$ on $\mathcal{H}$ is called a context $\Sigma$\smallskip

\begin{equation}  
   \Sigma
   =
   \left\{
      \hat{P}_A, \hat{P}_B, \dots
   \right\}
   \;\;\;\;   
\end{equation}
\smallskip

\noindent if the next requirements are satisfied:\smallskip

\begin{equation}  
   \hat{P}_A
   \hat{P}_B
   =
   \hat{P}_B
   \hat{P}_A
   =
   \hat{0}
   \;\;\;\;  ,
\end{equation}

\begin{equation}  
   \hat{P}_A
   +
   \hat{P}_B
   +
   \dots
   =
   \sum_{\hat{P} \in \Sigma}
   \hat{P}
   =
   \hat{1}
   \;\;\;\;  .
\end{equation}
\smallskip

\end{definition}

\noindent In view of the exact correspondence existing between the projection operator $\hat{P}$ and the proposition $P$ represented by the closed linear subspace $\mathrm{ran}(\hat{P})$, one may also refer to the context $\Sigma$ as the set of the compatible propositions:\smallskip

\begin{equation}  
   \Sigma
   =
   \left\{
      \hat{P}_A, \hat{P}_B, \dots
   \right\}
   \;
   \iff
   \;
   \Sigma
   =
   \left\{
      P_A, P_B, \dots
   \right\}
   \;\;\;\;  .
\end{equation}
\smallskip

\begin{definition} 
A subspace $\mathcal{P} \subseteq \mathcal{H}$ is called an invariant subspace under the projection operator $\hat{P}$ on $\mathcal{H}$ if\smallskip

\begin{equation}  
   \hat{P}
   \textnormal{:}
   \mkern10mu
   \mathcal{P}
   \to
   \mathcal{P}
   \;\;\;\;  .
\end{equation}
\smallskip

\end{definition}

\noindent This means that the image of every vector $|\Psi\rangle$ in $\mathcal{P}$ under $\hat{P}$ remains within $\mathcal{P}$ which can be written as\smallskip

\begin{equation}  
   \hat{P} \mathcal{P}
   =
   \left\{
      |\Psi\rangle \in \mathcal{P}
      \textnormal{:}
      \mkern10mu
      \hat{P} |\Psi\rangle
   \right\}
   \subseteq
   \mathcal{P}
   \;\;\;\;  .
\end{equation}
\smallskip

\noindent As it can be readily seen, $\mathcal{P} = \mathrm{ran}(\hat{P})$ and $\mathcal{P} = \mathrm{ker}(\hat{P})$, in addition to $\mathcal{P} = \{0\}$ and $\mathcal{P} = \mathcal{H}$.\\

\begin{definition} 
The set of all the invariant subspaces $\mathcal{P}$ of the Hilbert space $\mathcal{H}$ invariant under the projection operator $\hat{P}$ can be determined by:\smallskip

\begin{equation}  
   \mathcal{L}(\hat{P})
   =
   \left\{
      \mathcal{P} \subseteq \mathcal{H}
      \textnormal{:}
      \mkern10mu
      \hat{P} \mathcal{P}
      \subseteq
      \mathcal{P}
   \right\}
   \;\;\;\;  .
\end{equation}
\smallskip

\end{definition}

\noindent Consider the set of the invariant subspaces $\mathcal{L}(\Sigma)$ \textit{invariant under every projection operator} from the context $\Sigma$:\smallskip

\begin{equation}  
   \mathcal{L}(\Sigma)
   =
   \mathcal{L}(\hat{P}_A)
   \cap
   \mathcal{L}(\hat{P}_B)
   \cap
   \dots
   =
   \bigcap_{\hat{P} \in \Sigma}
   \mathcal{L}(\hat{P})
   \;\;\;\;  .
\end{equation}
\smallskip

\noindent The elements of this set form a complete lattice called \textit{the invariant-subspace lattice of the context} \cite{Radjavi}. The lattice operations on $\mathcal{L}(\Sigma)$ are defined in an ordinary way: For instance, \textit{the meet} $\wedge$ and \textit{the join} $\vee$ are defined by\smallskip

\begin{equation}  
   \mathcal{H}_A
   ,
   \mathcal{H}_B
   \in
   \mathcal{L}(\Sigma)
   \;\;
   \implies
   \;\;
   \left\{
      \begin{array}{l}
         \mathcal{H}_A
         \wedge
         \mathcal{H}_B
         =
         \mathcal{H}_A
         \cap
         \mathcal{H}_B
         \in
         \mathcal{L}(\Sigma)
         \\ 
         \mathcal{H}_A
         \vee
         \mathcal{H}_B
         =
         \left(
            (\mathcal{H}_A)^{\perp}
            \cap
            (\mathcal{H}_B)^{\perp}
         \right)^{\perp}
         \in
         \mathcal{L}(\Sigma)
      \end{array}
   \right.
   \;\;\;\;  .
\end{equation}
\smallskip

\noindent It is straightforward to verify that each invariant-subspace lattice $\mathcal{L}(\Sigma)$ contains only mutually commuting subspaces (corresponding to mutually commutable projection operators), which means that each $\mathcal{L}(\Sigma)$ is a Boolean algebra.\\

\begin{definition} 
The set $\{\mathcal{L}(\Sigma)\}_{\Sigma \in \mathcal{O}} \equiv \{\Sigma \!\in\! \mathcal{O}\!: \mathcal{L}(\Sigma)\}$ is the collection of the invariant-subspace lattices that is in one-to-one correspondence with the set of all the contexts $\mathcal{O}$ associated with the quantum system.

\end{definition}

\noindent If all the lattices from $\{\mathcal{L}(\Sigma)\}_{\Sigma \in \mathcal{O}}$ are pasted (or stitched) together at their common elements -- which are (aside from identical elements) the trivial subspaces $\{0\}$ and $\mathcal{H}$ -- then the resulted logic will be the Hilbert lattice $\mathcal{L}(\mathcal{H})$ \cite{Svozil08}:\smallskip

\begin{equation} \label{PAST} 
   \mathcal{L}(\mathcal{H})
   =
   \bigcup_{\Sigma \in \mathcal{O}}
      \mathcal{L}({\Sigma})
   \;\;\;\;  ,
\end{equation}
\smallskip

\noindent where $\cup$ denotes the set-theoretic union carried out simultaneously on elements of $\{\mathcal{L}(\Sigma)\}_{\Sigma \in \mathcal{O}}$. In this sense, \textit{the Hilbert lattice $\mathcal{L}(\mathcal{H})$ is the union of the collection $\{\mathcal{L}(\Sigma)\}_{\Sigma \in \mathcal{O}}$}.\\

\noindent Providing the set of all the contexts $\mathcal{O}$ form a continuum, the Hilbert lattice can also be called a continuum of pasting of the invariant-subspace lattices (Boolean algebras) $\mathcal{L}(\Sigma) \in \{\mathcal{L}(\Sigma)\}_{\Sigma \in \mathcal{O}}$.\\

\section{Logic account of the EPR paradox and its resolutions} \label{ch.3} 

\noindent The EPR paradox is a striking and influential thought experiment intended to defy the prediction of quantum mechanics that it is impossible to know both the position and the momentum of a quantum particle \cite{EPR}. In Bohm's formulation of this paradox \cite{Bohm}, the challenge is extended to the prediction that measurements of spin along the different axes are incompatible. The EPR paradox is still a centerpiece in the ongoing debates over the interpretation of quantum theory.\\

\noindent Let’s consider \textit{the logic account} of Bohm's formulation of the EPR paradox, that is, the description of the paradox based on experimentally verifiable propositions and their truth-values. The reason of this description is that it allows one to analyze the paradox using exclusively logic that is tied to the Hilbert space formalism of quantum mechanics, with no philosophical discussion on how to interpret the mathematical formulation of quantum mechanics.\\

\noindent Imagine a system with a pair of nonidentical one-half spin particles (e.g., an electron and a positron) which are prepared in a singlet state (i.e., a state with total spin angular momentum 0). Suppose that after being prepared, the particles 1 and 2 travel away from each other in a region of zero magnetic field where by means of Stern-Gerlach magnets corresponding observers 1 and 2 measure the spin of the matching particle.\\

\noindent When the observer $N \in \{1,2\}$ measures the spin of the particle $N$ along, say the $z$-axis, the observer verifies either the proposition ``The spin of the particle $N$ along the $z$-axis is $+\frac{\hbar}{2}\,$'' (abbreviated to the symbol $P_{Nz+}$) or the proposition ``The spin of the particle $N$ along the $z$-axis is $-\frac{\hbar}{2}\,$'' (abbreviated to the symbol $P_{Nz-}$) that are represented by the subspaces $\mathcal{H}_{1z+}$ and $\mathcal{H}_{1z-}$, respectively. Assume, the observer 1 verifies the proposition $P_{1z+}$, which can be described as\smallskip

\begin{equation}  
   {\big[\mkern-4.3mu\big[ 
      P_{1z+}
      \big(
         |\Psi_1\rangle \in \mathcal{H}_{1z+}
      \big)
   \big]\mkern-4.3mu\big]}_v
   =
   \mathrm{true}
   \;\;\;\;  ,
\end{equation}
\smallskip

\noindent where $|\Psi_1\rangle$ refers to the pure quantum state of the particle 1 (correspondingly, $|\Psi_2\rangle$ refers to the pure state of the particle 2).\\

\noindent Provided $\mathcal{H}_{1z-} = \mathcal{H}_{1z+}^{\perp}$ and so the subspaces $\mathcal{H}_{1z+}$ and $\mathcal{H}_{1z-}$ are \textit{incomparable with each other}, i.e., $\mathcal{H}_{1z+} \nleq \mathcal{H}_{1z-}$ and $\mathcal{H}_{1z-} \nleq \mathcal{H}_{1z+}$, this observer infers in accordance with (\ref{FALS}):\smallskip

\begin{equation}  
   {\big[\mkern-4.3mu\big[ 
      P_{1z-}
      \big(
         |\Psi_1\rangle \in \mathcal{H}_{1z+}
      \big)
   \big]\mkern-4.3mu\big]}_v
   =
   {\big[\mkern-4.3mu\big[ 
      P_{1z-}
      \big(
         |\Psi_1\rangle \notin \mathcal{H}_{1z-}
      \big)
   \big]\mkern-4.3mu\big]}_v
   =
   \mathrm{false}
   \;\;\;\;  .
\end{equation}
\smallskip

\noindent Moreover, in the structure of the Hilbert lattice $\mathcal{L}(\mathcal{H}_1)$, i.e., in the union of the invariant-subspace lattices of the contexts associated with the particle 1, the subspace $\mathcal{H}_{1z+}$ is also \textit{incomparable with both subspaces} $\mathcal{H}_{1x+}$ and $\mathcal{H}_{1x-}$ representing respectively the propositions ``The spin of the particle 1 along the $x$-axis is $+\frac{\hbar}{2}\,$'' and ``The spin of the particle 1 along the $x$-axis is $-\frac{\hbar}{2}\,$'' (abbreviated to the symbols $P_{1x+}$ and $P_{1x-}$ in that order). Consistent with (\ref{FALS}), this entails the following counter-factual truth-values of these propositions:\smallskip

\begin{equation} \label{X1} 
   {\big[\mkern-4.3mu\big[ 
     P_{1x\pm}
      \big(
         |\Psi_1\rangle \in \mathcal{H}_{1z+}
      \big)
   \big]\mkern-4.3mu\big]}_v
   =
   {\big[\mkern-4.3mu\big[ 
      P_{1x\pm}
      \big(
         |\Psi_1\rangle \notin \mathcal{H}_{1x\pm}
      \big)
   \big]\mkern-4.3mu\big]}_v
   =
   \mathrm{false}
   \;\;\;\;  .
\end{equation}
\smallskip

\noindent At this point, recall that the Hilbert space for two entangled (nonidentical) particles is the tensor product $\mathcal{H}_1 \!\otimes \mathcal{H}_2$ of the two Hilbert spaces $\mathcal{H}_1$ and $\mathcal{H}_2$ for the separate particles 1 and 2. By reason of the initial singlet state of the spins of these particles, when the particle 1 is in the state $|\Psi_1\rangle \in \mathcal{H}_{1z+}$, the composite system of the entangled particles must be in \textit{the product state}:\smallskip

\begin{equation} \label{PROD} 
   |\Psi_{12}\rangle
   \in
   \mathcal{Z}_{1}
   \equiv
   |\Psi_{1}\rangle \otimes |\Psi_{2}\rangle
   \in
   \mathcal{H}_{1z+} \otimes \mathcal{H}_{2z-}
   \;\;\;\;  .
\end{equation}
\smallskip

\noindent In accordance with the rule (\ref{RULE}), this implies\smallskip

\begin{equation} \label{Z12} 
   {\big[\mkern-4.3mu\big[ 
      \left(P_{1z+} \sqcap P_{2z-}\right)
      \left(
         |\Psi_{12}\rangle
         \in
         \mathcal{Z}_{1}
      \right)
   \big]\mkern-4.3mu\big]}_v
   =
   \mathrm{true}
   \;\;\;\;  ,
\end{equation}
\smallskip

\noindent where the abbreviation $(P_{1z+} \sqcap P_{2z-})$ stands for the proposition ``The spin of the particle 1 along the $z$-axis is $+\frac{\hbar}{2}\,$ \textit{and} the spin of the particle 2 along the $z$-axis is $-\frac{\hbar}{2}\,$''.\\

\noindent Because this proposition in the product state (\ref{PROD}) is equivalent to the disjunction of the propositions $P_{1z+}$ in the state $ |\Psi_{1}\rangle \in \mathcal{H}_{1z+}$ and $P_{2z-}$ in the state $|\Psi_{2}\rangle \in \mathcal{H}_{2z-}$ (see the exposition of this argument in \cite{Foulis, Griffiths}), i.e., in symbols,\smallskip

\begin{equation}  
   \left(P_{1z+} \sqcap P_{2z-}\right)
   \left(
      |\Psi_{12}\rangle
      \in
      \mathcal{Z}_{1}
   \right)
   =
   P_{1z+}
   \left(
      |\Psi_{1}\rangle
      \in
      \mathcal{H}_{1z+}
   \right)
   \sqcap
   P_{2z-}
   \left(
      |\Psi_{2}\rangle
      \in
      \mathcal{H}_{2z-}
   \right)
   \;\;\;\;  ,
\end{equation}
\smallskip

\noindent the observer 1 can calculate using (\ref{Z12}) the counter-factual truth-value of the proposition $P_{2z-}$:\smallskip

\begin{equation} \label{Z2} 
   {\big[\mkern-4.3mu\big[ 
      P_{2z-}
      \left(
         |\Psi_{1}\rangle
         \in
         \mathcal{H}_{1z+}
      \right)
   \big]\mkern-4.3mu\big]}_v
   =
   {\big[\mkern-4.3mu\big[ 
      P_{2z-}
      \left(
         |\Psi_{2}\rangle
         \in
         \mathcal{H}_{2z-}
      \right)
   \big]\mkern-4.3mu\big]}_v
   =
   \mathrm{true}
   \;\;\;\;  .
\end{equation}
\smallskip

\noindent Within the structure of the Hilbert lattice $\mathcal{L}(\mathcal{H}_1 \!\otimes \mathcal{H}_2)$ imposed on the closed linear subspaces of the tensor product $\mathcal{H}_1 \!\otimes \mathcal{H}_2$, the subspaces $\mathcal{Z}_{1} = \mathcal{H}_{1z+} \otimes \mathcal{H}_{2z-}$ and $\mathcal{Z}_{2} = \mathcal{H}_{1z-} \otimes \mathcal{H}_{2z+}$ are incomparable with $\mathcal{X}_{1} = \mathcal{H}_{1x+} \!\otimes \mathcal{H}_{2x-}$ and $\mathcal{X}_{2} = \mathcal{H}_{1x-} \!\otimes \mathcal{H}_{2x+}$, i.e., both subspaces that represent the propositions ``The spin of the particle 1 along the $x$-axis is $\pm\frac{\hbar}{2}\,$ \textit{and} the spin of the particle 2 along the $x$-axis is $\mp\frac{\hbar}{2}\,$'' abbreviated into $(P_{1x\pm} \sqcap P_{2x\mp})$. As an inference from this fact, the observer 1 obtains the counter-factual truth-values of $(P_{1x\pm} \sqcap P_{2x\mp})$ similar to ones in (\ref{X1}):\smallskip

\begin{equation} \label{X12} 
   {\big[\mkern-4.3mu\big[ 
     \left(P_{1x\pm} \sqcap P_{2x\mp}\right)
      \big(
         |\Psi_{12}\rangle \in \mathcal{Z}_{1}
      \big)
   \big]\mkern-4.3mu\big]}_v
   =
   {\big[\mkern-4.3mu\big[ 
     \left(P_{1x\pm} \sqcap P_{2x\mp}\right)
      \big(
         |\Psi_{12}\rangle \notin \mathcal{X}_{1,2}
      \big)
   \big]\mkern-4.3mu\big]}_v
   =
   \mathrm{false}
   \;\;\;\;  .
\end{equation}
\smallskip

\noindent On the other hand, since the propositions $(P_{1x\pm} \sqcap P_{2x\mp})$ in the product state (\ref{PROD}) are equivalent to the disjunctions of the propositions, i.e.,\smallskip

\begin{equation}  
   \left(P_{1x\pm} \sqcap P_{2x\mp}\right)
   \left(
      |\Psi_{12}\rangle
      \in
      \mathcal{Z}_{1}
   \right)
   =
   P_{1x\pm}
   \left(
      |\Psi_{1}\rangle
      \in
      \mathcal{H}_{1z+}
   \right)
   \sqcap
   P_{2x\mp}
   \left(
      |\Psi_{2}\rangle
      \in
      \mathcal{H}_{2z-}
   \right)
   \;\;\;\;  ,
\end{equation}
\smallskip

\noindent and, thanks to (\ref{X1}), $P_{1x\pm}$ have the definite counter-factual truth-values in the state $|\Psi_1\rangle \in \mathcal{H}_{1z+}$, the observer 1 finds:\smallskip

\begin{equation}  
   {\big[\mkern-4.3mu\big[
      P_{2x\mp}
      \left(
         |\Psi_{2}\rangle
         \in
         \mathcal{H}_{2z-}
      \right)
   \big]\mkern-4.3mu\big]}_v
   =
   \mathfrak{t}
   \in
   \{ \mathrm{true},\mathrm{false} \}
   \;\;\;\;  .
\end{equation}
\smallskip

\noindent Together with (\ref{Z2}) it may possibly imply that the quantum propositions $P_{2z-}$ and $P_{2x-}$, equally as $P_{2z-}$ and $P_{2x+}$, can be verified simultaneously, i.e.,\smallskip

\begin{equation} \label{CON} 
   {\big[\mkern-4.3mu\big[
      P_{2z-}
      \left(
         |\Psi_{2}\rangle
         \in
         \mathcal{H}_{2z-}
      \right)
      \sqcap
      P_{2x\mp}
      \left(
         |\Psi_{2}\rangle
         \in
         \mathcal{H}_{2z-}
      \right)
   \big]\mkern-4.3mu\big]}_v
   =
   \mathrm{true}
   \;\;\;\;  .
\end{equation}
\smallskip

\noindent This constitutes the EPR paradox: From apparently true premises (\ref{X1}) and (\ref{X12}), and by seemingly uncontroversial reasoning, the conclusion is drawn that contradicts quantum mechanics.\\

\noindent To avoid the EPR paradoxical conclusion (\ref{CON}), one may assume that \textit{the principle of locality} does not hold and, therefore,\smallskip

\begin{equation}  
   {\big[\mkern-4.3mu\big[
      P_{1z+}
      \left(
         |\Psi_{1}\rangle
         \in
         \mathcal{H}_{1z+}
      \right)
      \sqcap
      P_{2x\mp}
      \left(
         |\Psi_{2}\rangle
         \in
         \mathcal{H}_{2z-}
      \right)
   \big]\mkern-4.3mu\big]}_v
   =
   \mathrm{false}
   \;\;\;\;  .
\end{equation}
\smallskip

\noindent That is, the verification of the proposition $P_{1z+}$ makes false not only both $P_{1x-}$ and $P_{1x+}$, i.e., the propositions which relate to the same particle, but also both $P_{2x-}$ and $P_{2x+}$, the propositions relating to another particle situated at a distance away.\\

\noindent Alternatively, i.e., making the conclusion (\ref{CON}) incorrect but with that holding the principle of locality, one may deny propositional counter-factual definiteness in the case where the quantum propositions cannot be verified simultaneously, i.e., they are incompatible. Along these lines, one can maintain that the counter-factual truth-values (\ref{X1}) and (\ref{X12}) are not \textit{admissible} \cite{Abbott, Svozil17} because they violate the following condition:\smallskip

\begin{equation} \label{ADM} 
   \sum_{P \in \Sigma}
   {\big[\mkern-4.3mu\big[ P \big]\mkern-4.3mu\big]}_b
   =
   1
   \;\;\;\;  ,
\end{equation}
\smallskip

\noindent where $b$ stand for \textit{the bivaluation} \cite{Beziau}, i.e., the assignment of the values 1 and 0 (which denote true and false, respectively) to the propositions $P$ from the context $\Sigma$. Due to the violation of the stipulated condition for the context $\Sigma_{1x} = \{P_{1x+}, P_{1x-} \}$ in the state $|\Psi_1\rangle \in \mathcal{H}_{1z+}$, the propositions $P_{1x\pm}$ must be declared undefined in this state, i.e.,\smallskip

\begin{equation}  
   {\big[\mkern-4.3mu\big[ 
     P_{1x\pm}
      \big(
         |\Psi_1\rangle \in \mathcal{H}_{1z+}
      \big)
   \big]\mkern-4.3mu\big]}_b
   \neq
   \mathfrak{b}
   \in
   \{ 1,0 \}
   \;\;\;\;  .
\end{equation}
\smallskip

\noindent Likewise,\smallskip

\begin{equation}  
   {\big[\mkern-4.3mu\big[
      \left(P_{1x\pm} \sqcap P_{2x\mp}\right)
      \left(
         |\Psi_{12}\rangle
         \in
         \mathcal{Z}_{1}
      \right)
   \big]\mkern-4.3mu\big]}_b
   \neq
   \mathfrak{b}
   \in
   \{ 1,0 \}
   \;\;\;\;  .
\end{equation}
\smallskip

\section{Regarding the notion of admissibility}  

\noindent The admissibility condition (\ref{ADM}) stems from the following hypothesis:\smallskip

\begin{hypothesis} \label{H} 
Probabilities and expectations on Boolean algebras (or blocks) of an empirical logic are classical \textnormal{\cite{Svozil18}}.

\end{hypothesis}

\vspace*{2mm}

\noindent This means that as long as the bivaluation ${[\![ P ]\!]}_b = \mathfrak{b} \in \{ 1,0 \}$ can be interpreted as the dispersion-free probability measure, propositions from one and the same context $\Sigma$ must obey the Kolmogorov axioms \cite{Kolmogorov}.\\

\noindent In particular, the probability that the disjunction of the propositions $P_{1x+}$ and $P_{1x-}$ is verified, that is, $\Pr\mkern2mu[ P_{1x+} \sqcup P_{1x-} \text{ is true} \mkern2mu]$, must be equal to the sum of the probabilities of the propositions $P_{1x+}$ and $P_{1x-}$ being verified, i.e., $\Pr\mkern2mu[ P_{1x+} \text{ is true} \mkern2mu]$ and $\Pr\mkern2mu[ P_{1x-} \text{ is true} \mkern2mu]$. Since on each context $\Sigma$ the probabilities add to 1, the condition (\ref{ADM}) then follows:\smallskip

\begin{equation}  
   \Pr\big[ P_{1x+} \text{ is true} \mkern2.5mu\big]
   +
   \Pr\big[ P_{1x-} \text{ is true} \mkern2.5mu\big]
   =
   {\big[\mkern-4.3mu\big[ P_{1x+} \big]\mkern-4.3mu\big]}_b
   +
   {\big[\mkern-4.3mu\big[ P_{1x-} \big]\mkern-4.3mu\big]}_b
   =
   1
   \;\;\;\;  .
\end{equation}
\smallskip

\noindent Nevertheless, even supposing that the hypothesis \ref{H} is correct, the question remains: How does a probability concept appear in an empirical logic? For, without Kolmogorov’s axiomatic system one cannot ground the requirement of admissibility for the quantum propositions.\\

\noindent Of course, one may simply assume that a probability space exists for the quantum propositions and then introduce the admissibility requirement. But in that case, the rejection of truth-values of the counterfactually definite kind, such as (\ref{X1}) and (\ref{X12}), turns out to be some additional postulate unwarranted by the Hilbert space formalism.\\

\noindent Hence, replacing this postulate by a derivation resulting from the Hilbert space formalism would be a greatly desirable as a means to resolve the EPR paradox.\\

\section{The structure of invariant-subspace lattices free from pasting}  

\noindent When there is no gathering into the Hilbert lattice $\mathcal{L}(\mathcal{H})$ of the invariant-subspace lattices $\mathcal{L}(\Sigma)$ of the contexts $\Sigma$ associated with the quantum system, the nonidentical subspaces $\mathcal{H}^{\prime}$ and $\mathcal{H}^{\prime\prime}$ belonging to the different lattices $\mathcal{L}(\Sigma^{\prime})$ and $\mathcal{L}(\Sigma^{\prime\prime})$ cannot be in one (partially ordered) set.\\

\noindent This means, in particular, that for the subspaces $\mathcal{H}_{Nz\pm} \in \mathcal{L}(\Sigma_{Nz})$ where $\Sigma_{Nz} = \{P_{Nz+}, P_{Nz-} \}$ and the subspaces $\mathcal{H}_{Nx\pm} \in \mathcal{L}(\Sigma_{Nx})$ where $\Sigma_{Nx} = \{P_{Nx+}, P_{Nx-} \}$, ordering relation has no meaning at all since $\{ \mathcal{H}_{Nz\pm}, \mathcal{H}_{Nx\pm} \}$ are not subsets of any set $\mathcal{L}(\Sigma_{Nq})$ from the collection\smallskip

\begin{equation}  
   \Big\{
      \mathcal{L}(\Sigma_{Nq})
   \Big\}_{q \in \mathcal{R}^3}
   \equiv
   \bigg\{
      q \in \mathcal{R}^3
      \!:
      \;
      \Big\{
         \{0\}
         \,
         ,
         \,
         \mathcal{H}_{Nq+}
         \,
         ,
         \,
         \mathcal{H}_{Nq-}
         \,
         ,
         \,
         \mathbb{C}^2
      \Big\}
   \bigg\}
   \;\;\;\;  ,
\end{equation}
\smallskip

\noindent where $q$ represents an arbitrary axis. In other words, within this collection, the subspaces $\mathcal{H}_{Nz\pm}$ and $\mathcal{H}_{Nx\pm}$ are \textit{neither comparable nor incomparable}. This can be expressed equivalently by stating that the lattice-theoretic meet cannot be defined on such elements, which in symbols can be written down as\smallskip

\begin{equation}  
   \{ \mathcal{H}_{Nz\pm}, \mathcal{H}_{Nx\pm} \}
   \mkern5mu\cancel{\subseteq}\mkern5mu
   \mathcal{L}(\Sigma_{Nq})
   \in
   \Big\{
      \mathcal{L}(\Sigma_{Nq})
   \Big\}_{q \in \mathcal{R}^3}
   :
   \;\;
   \mathcal{H}_{Nz\pm}
   \mkern5mu\cancel{\wedge}\mkern5mu
   \mathcal{H}_{Nx\pm}
   \;\;\;\;  ,
\end{equation}
\smallskip

\noindent where the diagonal strikeout over $\wedge$ signifies that under the condition of no pasting into $\mathcal{L}(\mathcal{H}_N)$ the meet operation $\wedge$ cannot be defined for the indicated subspaces (recall that the meet is defined as an operation on pairs of elements from one partially ordered set).\\

\noindent But if the subspaces $\mathcal{H}_{1z\pm}$ are neither comparable nor incomparable with the subspaces $\mathcal{H}_{1x\pm}$, the counter-factual truth-values of the quantum propositions $P_{1x\pm}$ in the states $|\Psi_1\rangle \in \mathcal{H}_{1z\pm}$ are meaningless, i.e., without the truth values. This can be expressed by\smallskip

\begin{equation}  
   \{ \mathcal{H}_{1z\pm}, \mathcal{H}_{1x\pm} \}
   \mkern5mu\cancel{\subseteq}\mkern5mu
   \mathcal{L}(\Sigma_{1q})
   \in
   \Big\{
      \mathcal{L}(\Sigma_{1q})
   \Big\}_{q \in \mathcal{R}^3}
   :
   \;\;
   {\big[\mkern-4.3mu\big[ 
     P_{1x\pm}
      \big(
         |\Psi_1\rangle \in \mathcal{H}_{1z+}
      \big)
   \big]\mkern-4.3mu\big]}_b
   =
   \frac{0}{0}
   \;\;\;\;  ,
\end{equation}
\smallskip

\noindent where $\frac{0}{0}$ denotes an indeterminate value. Similarly,\smallskip

\begin{equation}  
   \{
      \mathcal{H}_{1z\pm}
      \!\otimes\!
      \mathcal{H}_{2z\mp}
      \mkern4mu
      \!
      ,
      \mkern4.5mu      
      \mathcal{H}_{1x\pm}
      \!\otimes\!
      \mathcal{H}_{2x\mp}
    \}
   \mkern5mu\cancel{\subseteq}\mkern5mu
   \mathcal{L}(\Sigma)
   \!\in\!
   \{\mathcal{L}(\Sigma)\}_{\Sigma \in \mathcal{O}}
   :
   \;
   {\big[\mkern-4.3mu\big[
      \left(P_{1x\pm} \sqcap P_{2x\mp}\right)
      \left(
         |\Psi_{12}\rangle
         \in
         \mathcal{Z}_{1}
      \right)
   \big]\mkern-4.3mu\big]}_b
   =
   \frac{0}{0}
   \;\;\;\;  .
\end{equation}
\smallskip

\noindent Speaking in general terms, in the case that the quantum propositions $P$ and $Q$ are affiliated with the different \textit{non-intertwined} contexts $\Sigma_P$ and $\Sigma_Q$ (meaning that $\Sigma_P$ and $\Sigma_Q$ do not share common elements \cite{Svozil17}) and, consequently, the subspaces $\mathcal{H}_P$ and $\mathcal{H}_Q$ representing $P$ and $Q$ are from the different non-intertwined lattices, namely, $\mathcal{H}_P \subseteq \mathcal{L}(\Sigma_P)$ and $\mathcal{H}_Q \subseteq \mathcal{L}(\Sigma_Q)$, propositional counter-factual definiteness cannot appear. Hence, in the structure of the collection $\{\mathcal{L}(\Sigma_P), \mathcal{L}(\Sigma_Q), \dots \}$ with no pasting, the hypothesis of propositional counter-factual definiteness (\ref{HYP}) does not hold true:\smallskip

\begin{equation}  
   \exists
   \left\{
      \mathcal{H}_Q
      ,
      \mathcal{H}_P
   \right\}
   \mkern5mu\cancel{\subseteq}\mkern5mu
   \mathcal{L}(\Sigma)
   \in
   \left\{\mathcal{L}(\Sigma_P), \mathcal{L}(\Sigma_Q), \dots \right\}
   \textnormal{:}
   \;\;
   {\big[\mkern-4.3mu\big[ 
     P
      \big(
         |\Psi\rangle \in \mathcal{H}_Q
      \big)
   \big]\mkern-4.3mu\big]}_b
   =
   \frac{0}{0}
   \;\;\;\;  .
\end{equation}
\smallskip

\noindent More to this point, consider the sequence $\mathcal{S}$ of the quantum propositions relating to the system of two entangled spin-half particles examined in the section \ref{ch.3}:\smallskip

\begin{equation}  
   \mathcal{S}
   =
   \Big(
      \left(P_{1z+} \!\sqcap\! P_{2z-}\right)
      ,
      \,
      \left(P_{1z-} \!\sqcap\! P_{2z+}\right)
      ,
      \,
      \left(P_{1x+} \!\sqcap\! P_{2x-}\right)
      ,
      \,
      \left(P_{1x-} \!\sqcap\! P_{2x+}\right)
   \Big)
   \;\;\;\;  .
\end{equation}
\smallskip

\noindent Within the structure $\{\mathcal{L}(\Sigma)\}_{\Sigma \in \mathcal{O}}$ free from pasting, \textit{the population of the truth-values} for $\mathcal{S}$ in the states $|\Psi_{12}\rangle \in \mathcal{Z}_{1,2}$ are tuples $\mathcal{T}_1$ and $\mathcal{T}_2$ that do not include counterfactually definite truth-values for incompatible propositions:\smallskip

\begin{equation}  
   \mathcal{T}_1
   =
   \Big(
      1
      ,
      \,
      0
      ,
      \,
      \frac{0}{0}
      \,
      ,
      \,
      \frac{0}{0}
   \Big)
   \;\;\;\;  ,
\end{equation}

\begin{equation}  
   \mathcal{T}_2
   =
   \Big(
      0
      ,
      \,
      1
      ,
      \,
      \frac{0}{0}
      \,
      ,
      \,
      \frac{0}{0}
   \Big)
   \;\;\;\;  .
\end{equation}
\smallskip

\section{Conclusion remarks}  

\noindent As it is shown through the logic account of the EPR paradox, to avoid the paradoxical inference (\ref{CON}), one can deny (apart from the completeness of quantum mechanics) either the principle of locality or the hypothesis of propositional counter-factual definiteness.\\

\noindent But then again, in the union of the collection $\{\mathcal{L}(\Sigma)\}_{\Sigma \in \mathcal{O}}$, the counter-factual truth-values are always definite. Indeed, because in this union, a pair of closed linear subspaces $\mathcal{H}_Q$ and $\mathcal{H}_P$ is either comparable or incomparable, the hypothesis of propositional counter-factual definiteness holds true. So, if one is unwilling to abandon the principle of locality in the union of $\{\mathcal{L}(\Sigma)\}_{\Sigma \in \mathcal{O}}$, one is compelled to introduce one more hypothesis – namely, the requirement of admissibility – just to make void counterfactually definite truth-values of incompatible propositions holding in this union.\\

\noindent By contrast, in the collection $\{\mathcal{L}(\Sigma)\}_{\Sigma \in \mathcal{O}}$ free from pasting, the closed linear subspaces representing incompatible propositions are neither comparable nor incomparable with each other. As a result, propositional counter-factual definiteness cannot exist for such propositions.\\

\noindent One can conclude from this that the Hilbert lattice $\mathcal{L}(\mathcal{H})$, i.e., the union of the collection $\{\mathcal{L}(\Sigma)\}_{\Sigma \in \mathcal{O}}$, is an unnecessary assumption in an empiric logic.\\

\section*{Acknowledgment}
\noindent The author would like to express his gratitude to the anonymous referee for the positive feedback and valuable insights.\\

\bibliographystyle{References}

\end{document}